\documentclass[aps,prb, twocolumn]{revtex4-1}

\usepackage{fancyhdr}
\usepackage{amsmath}
\usepackage{graphicx}
\usepackage{lipsum}

\begin{document}

\title{Valley blockade in a silicon double quantum dot}
\author{Justin K. Perron}
\email{jperron@csusm.edu}
\affiliation{California State University San Marcos, Department of Physics, California 92096}
\affiliation{Joint Quantum Institute, National Institute of Standards and Technology, Gaithersburg, Maryland 20899, USA}
\author{Michael J. Gullans}
\affiliation{Joint Quantum Institute, National Institute of Standards and Technology, Gaithersburg, Maryland 20899, USA}
\affiliation{Joint Center for Quantum Information and Computer Science, University of Maryland, College Park, Maryland 20742, USA}
\author{Jacob M. Taylor}
\affiliation{Joint Quantum Institute, National Institute of Standards and Technology, Gaithersburg, Maryland 20899, USA}
\affiliation{Joint Center for Quantum Information and Computer Science, University of Maryland, College Park, Maryland 20742, USA}
\author{M. D. Stewart, Jr.}
\author{Neil M. Zimmerman}
\email{Neil.Zimmerman@nist.gov, Tel: (301) 975-5887}
\affiliation{National Institute of Standards and Technology, Gaithersburg, Maryland 20899, USA}

\begin{abstract}
Electrical transport in double quantum dots (DQDs) illuminates many interesting features of the dots' carrier states. Recent advances in silicon quantum information technologies have renewed interest in the valley states of electrons confined in silicon.  Here we show measurements of DC transport through a mesa-etched silicon double quantum dot.  Comparing bias triangles (i.e., regions of allowed current in DQDs) at positive and negative bias voltages we find a systematic asymmetry in the size of the bias triangles at the two bias polarities.  Asymmetries of this nature are associated with blocked tunneling events due to the occupation of a metastable state. Several features of our data lead us to conclude that the  states involved are not simple spin states. Rather, we develop a model based on selective filling of valley states in the DQD that is consistent with all of the qualitative features of our data.
\end{abstract}

\maketitle

The conduction band of an indirect semiconductor has multiple degenerate minima. Silicon, for example, has six equivalent minima (or valleys) at wave vectors 85\% of the way to the zone boundary in the equivalent [100] directions. This means conduction electrons have an additional degree of freedom when compared to those in direct gap semiconductors (with conduction band minima centered at $k=0$).  Although some implications of this valley degree of freedom were measured as far back as 1966 in Shubnikov-de Haas oscillations\cite{Fowler66}, there has been a recent spike in the amount of work focusing on the valley properties of conduction electrons in silicon.  This includes measurements of valley splittings in different Si-based quantum dots\cite{Kawakami14, Borselli11B,Lim11,Yang13,Hao14,Fuechsle10,Takashina06}. Much of the impetus behind this interest is due to recent developments in solid state quantum computation. These developments have highlighted several ways in which the valley state of conduction electrons can influence the quantum behaviour of confined electrons.  This includes 
valley induced oscillations of the exchange interaction over atomic length scales\cite{Koiller01}, and spin relaxation hot spots\cite{Yang13}. Valley states are also believed to influence the voltage induced g-factor shift of a quantum dot spin\cite{Veldhorst15} which enables spin qubit addressability\cite{Veldhorst14}. Furthermore, when creating an electron spin qubit one needs two isolated spin states; thus it is beneficial for the lowest lying valley states be separated by an energy larger than the Zeeman splitting of the spin states. 

In this paper, we report transport measurements of a silicon double quantum dot that reveal a rectification effect between bias voltage polarities. To explain our data we propose a model involving the valley degree of freedom and a substantial difference in the electron filling of the two quantum dots. The model suggests that this type of blockade could be used to probe several aspects of valley physics similar to how PSB has been used to probe solid state spin physics.

\begin{figure}
	\centering
	\includegraphics[width = 0.9\columnwidth]{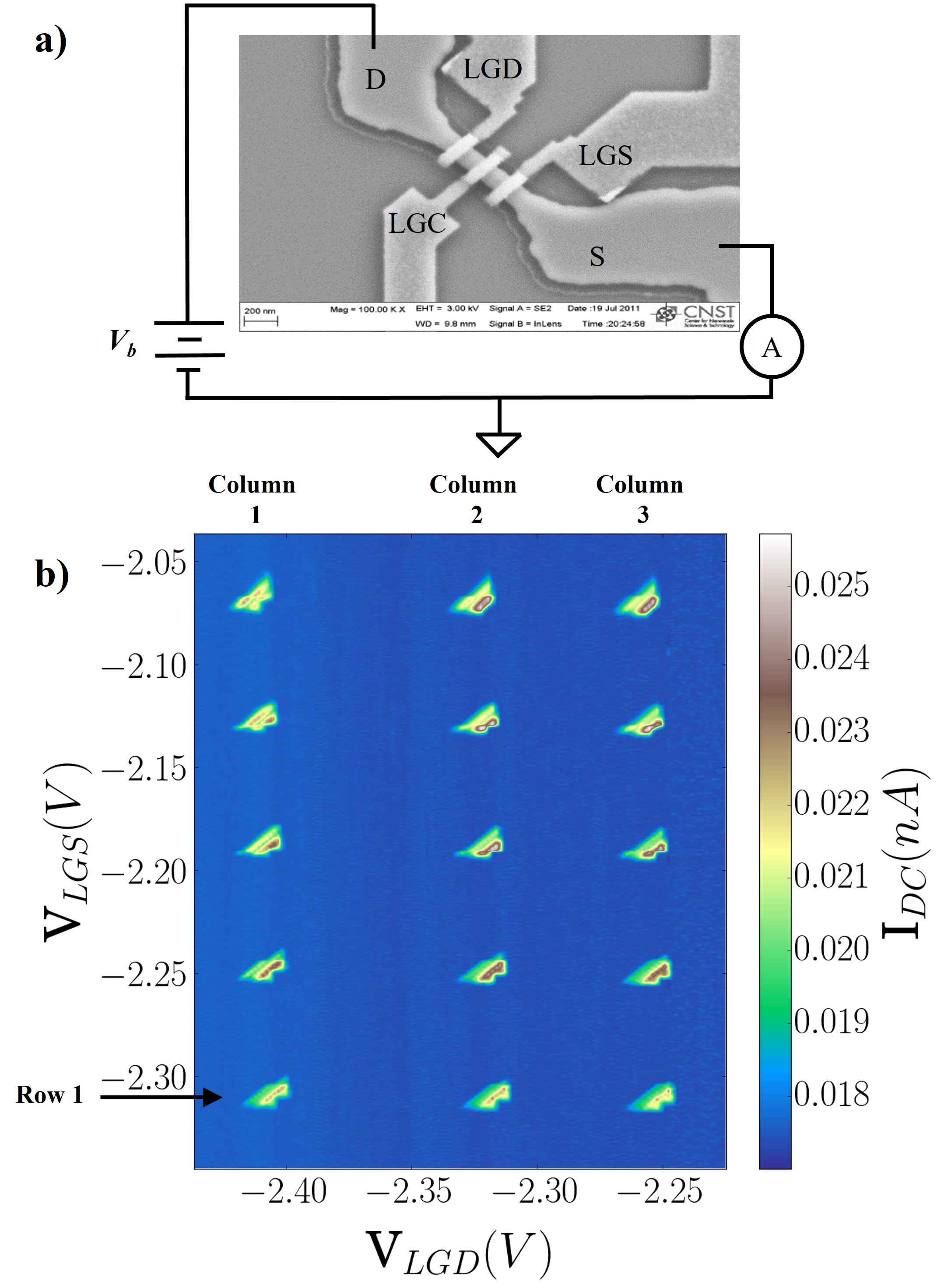}   
	\caption{{\bf{Charge Stability Diagram:}} {\bf{a)}} Scanning electron micrograph of a device similar to the one measured. Current is measured at the source lead with the bias voltage applied to the drain. Not shown is a global upper-gate used to turn on conduction in the silicon. {\bf{b)}} DC transport through our double quantum dot measured with $V_b$~=~1.5~mV, an upper gate voltage of $V_{\rm{UG}}$~=~3.5~V, and a center lower gate voltage of $V_{\rm{LGC}}~=~-1.733~{\rm{V}}$. This measurement spans five charge transitions on the source side dot and three on the drain side dot.  At lower $V_{\rm{LGS}}$ and $V_{\rm{LGD}}$ values our noise floor of $\pm 0.1$~pA prevents measurement of bias triangles at very low current values, while at higher values the bias triangles are not very well formed as cotunneling effects begin to become evident. All voltages are referenced to ground and the apparent background current is a result of an offset in the current preamplifier. \label{grid}}
\end{figure}

Our device [see figure~\ref{grid}a)] consists of a mesa-etched silicon nanowire formed from a (100) silicon-on-insulator substrate\cite{Koppinen13}. A SiO$_2$ dielectric layer separates three polysilicon finger gates from the nanowire. These gates, spaced 40~nm edge-to-edge,  conformally coat the nanowire and are used to electrostatically create tunnel barriers.  A second SiO$_2$ dielectric layer electrically isolates the finger gates from a polysilicon global upper gate which is used to turn on conduction in the device. Far from the active device area shown, ohmic contacts are formed on degenerately doped regions of the mesa-etched silicon. In addition to forming the barriers between the quantum dots and the source/drain leads, the outermost finger gates also serve as plunger gates, raising and lowering the chemical potentials of the quantum dots.

Figure~\ref{grid}b) shows DC transport measured in our device as a function of the voltages on the outer two finger gates, $V_{\rm{LGD}}$ and $V_{\rm{LGS}}$. All data presented are taken in a dilution refrigerator at a nominal base temperature of 45~mK. The measurement results in a honeycomb stability diagram where each cell of the honeycomb corresponds to a constant number of electrons on each dot\cite{VDW02}. At the corners of each hexagonal cell are regions, called bias triangles, where a tunneling current is energetically favorable.  Since the applied source-drain bias voltage, $V_b$, determines the energy window between the Fermi levels of the two leads, one expects the size of the bias triangles to be proportional to\cite{VDW02}  $|V_b|$, and that the polarity of the bias should not change the size of the triangles. 

\begin{figure}
	\centering
	\includegraphics[width = 0.9\columnwidth]{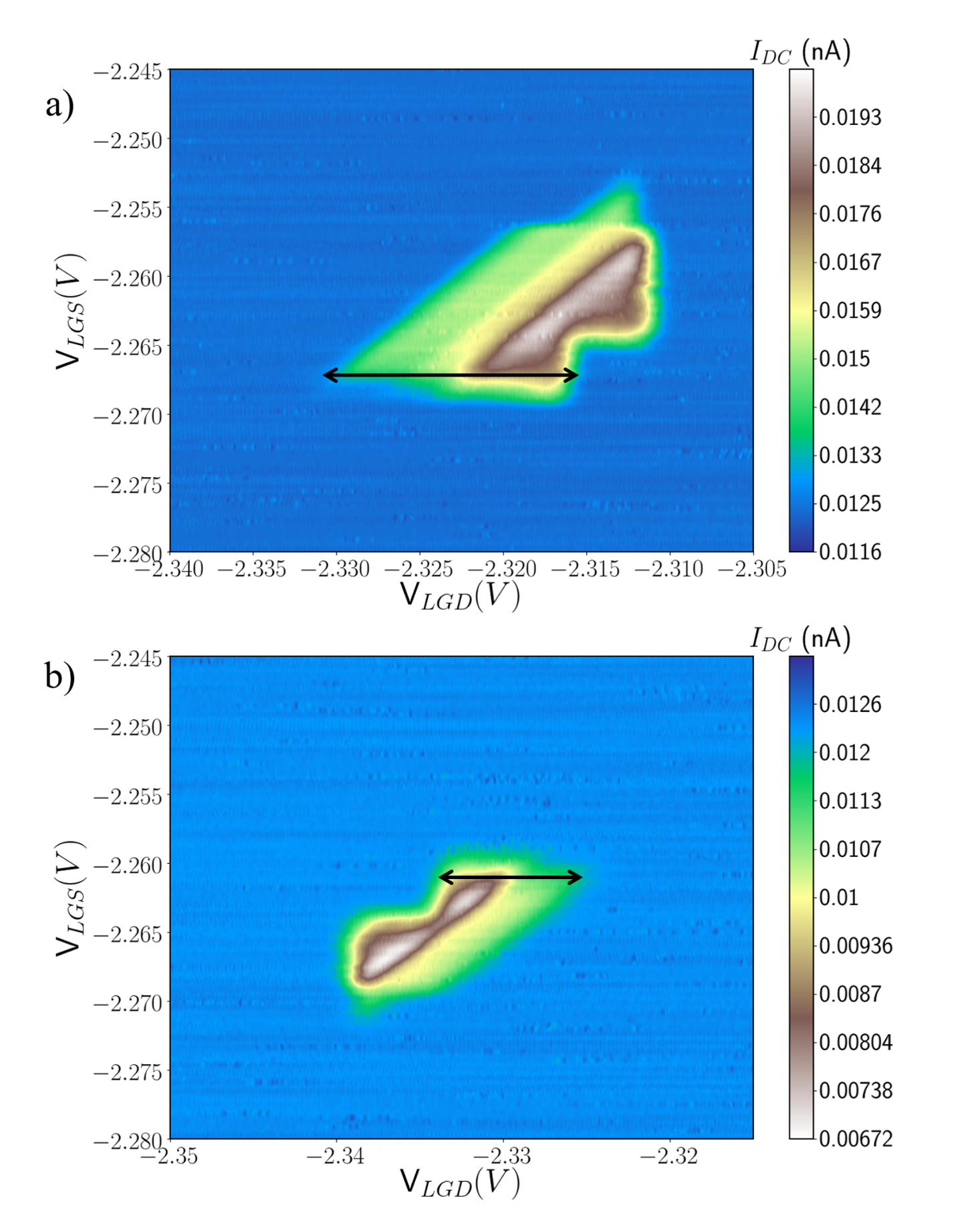}   
	\caption{{\bf{Size Asymmetry}} A finer measurement of the bias triangles from column~2 and row~2 in figure~\ref{grid}. {\bf{a)}} Measured with $V_b$~=~1.5~mV, {\bf{b)}} measured with $V_b$~=~-1.5~mV. The size asymmetry of the triangles can be quantified by $\Delta E_{\rm{open}}$ calculated using the $V_{\rm{open}}$ values indicated by the black arrows. Similar asymmetries were observed on the majority of charge transitions pictured in figure~\ref{grid}.\label{triangles}}
\end{figure}

Data taken on a finer scale focusing on a single set of bias triangles is shown in figure~\ref{triangles}, a) taken with $V_b~=~1.5~{\rm{mV}}$ and b) $V_b~=~-1.5~{\rm{mV}}$. There is a clear asymmetry in the size of the triangles for the two polarities contradicting the expectation that the polarity of $V_b$ would not affect the triangle size.

\begin{figure}
	\centering
	\includegraphics[width = 0.9\columnwidth]{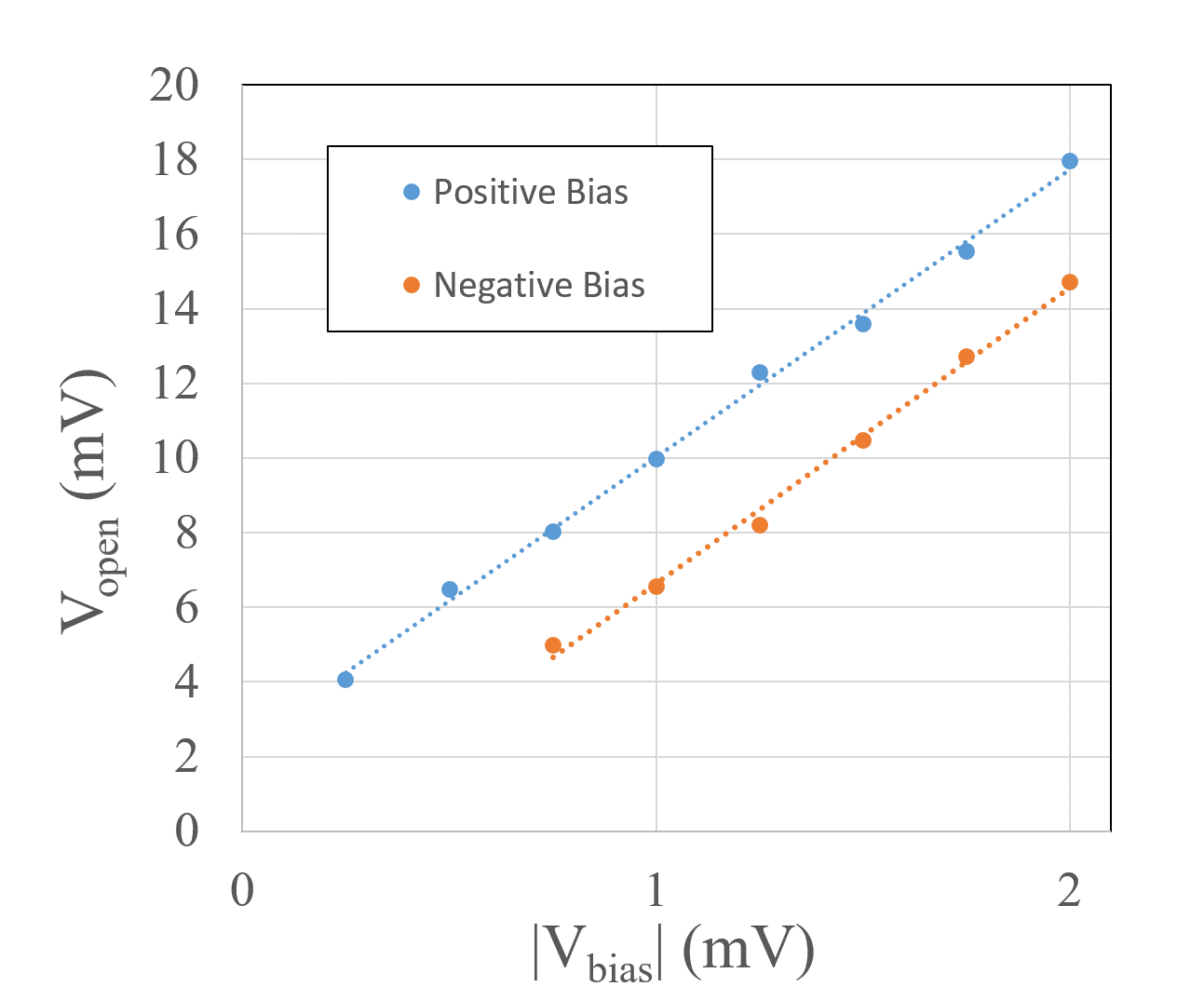}   
	\caption{{\bf{Triangle size vs. bias voltage:}} Bias triangle size increases linearly with the magnitude of the bias voltage $V_b$. The slopes from the $V_b >0$ and $V_b <0$ polarities are 7.7 and 7.9 respectively. The asymmetry observed in figure~\ref{triangles} is consistent at all $|V_b|$ investigated.\label{Bias}}
\end{figure}

To quantify the size of the triangles we use the width of the triangles in $V_{LGD}$ which we call $V_{open}$ (indicated by the black arrows in figure~\ref{triangles}). As shown in figure~\ref{Bias}, The size of the triangles does scale linearly with $|V_b|$ for both positive and negative biases\footnote{One expects the trend in this figure to intercept the origin which it does not. We were unable to confirm a specific source of this.} . Furthermore, the size asymmetry between the two bias polarities is essentially constant across a wide range of $|V_b|$. $V_{open}$ corresponds to the change in $V_{\rm{LGD}}$ that shifts the chemical potential of the drain dot across the energy window where conduction is allowed.  In a typical non-blockaded situation this window corresponds to the bias window $e|V_b|$.  In all our measurements showing a size asymmetry the positively biased triangles were larger than the negatively biased triangles. Thus, we assume the positive biased triangles correspond to a non-blockaded situation allowing us to define a lever arm in the same manner described in \cite{VDW02} 

\begin{equation}
\alpha = \frac{e\left|V_b\right|}{V_{open}|_{V_b>0}},
\end{equation}

\noindent with $e$ being the electron charge.  This $\alpha$ converts $V_{open}$ to an energy $E_{open}=\alpha V_{open}$.  The asymmetry can then be quantified in units of energy by 

\begin{equation}
\Delta E_{open} = e\left|V_b\right| - \alpha V_{open}|_{V_b<0}.
\end{equation}

\begin{figure*}
	\centering
	\includegraphics[width = 0.9\textwidth]{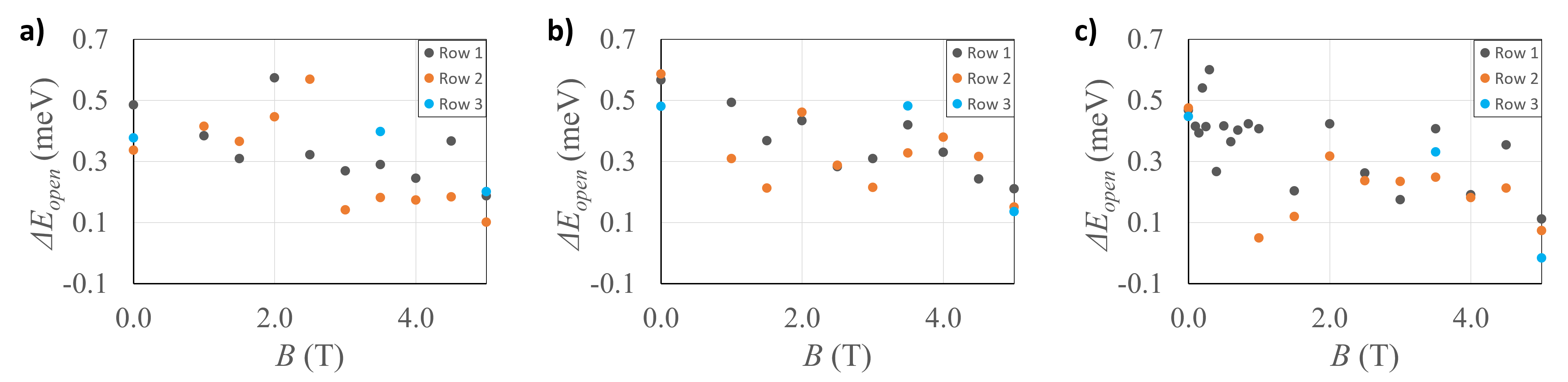}   
	\caption{{\bf{Magnetic Field Dependence:}} Size asymmetry as a function of applied magnetic field for the first three rows of transitions shown in figure~\ref{grid}. a) Column 1, b) column 2, and c) column 3. Although the magnetic field clearly changes the magnitude of the size asymmetry, there lacks a clean systematic dependence as one would expect from spin states. \label{Bfield}}
\end{figure*}

\begin{figure}
	\centering
	\includegraphics[width = 0.9\columnwidth]{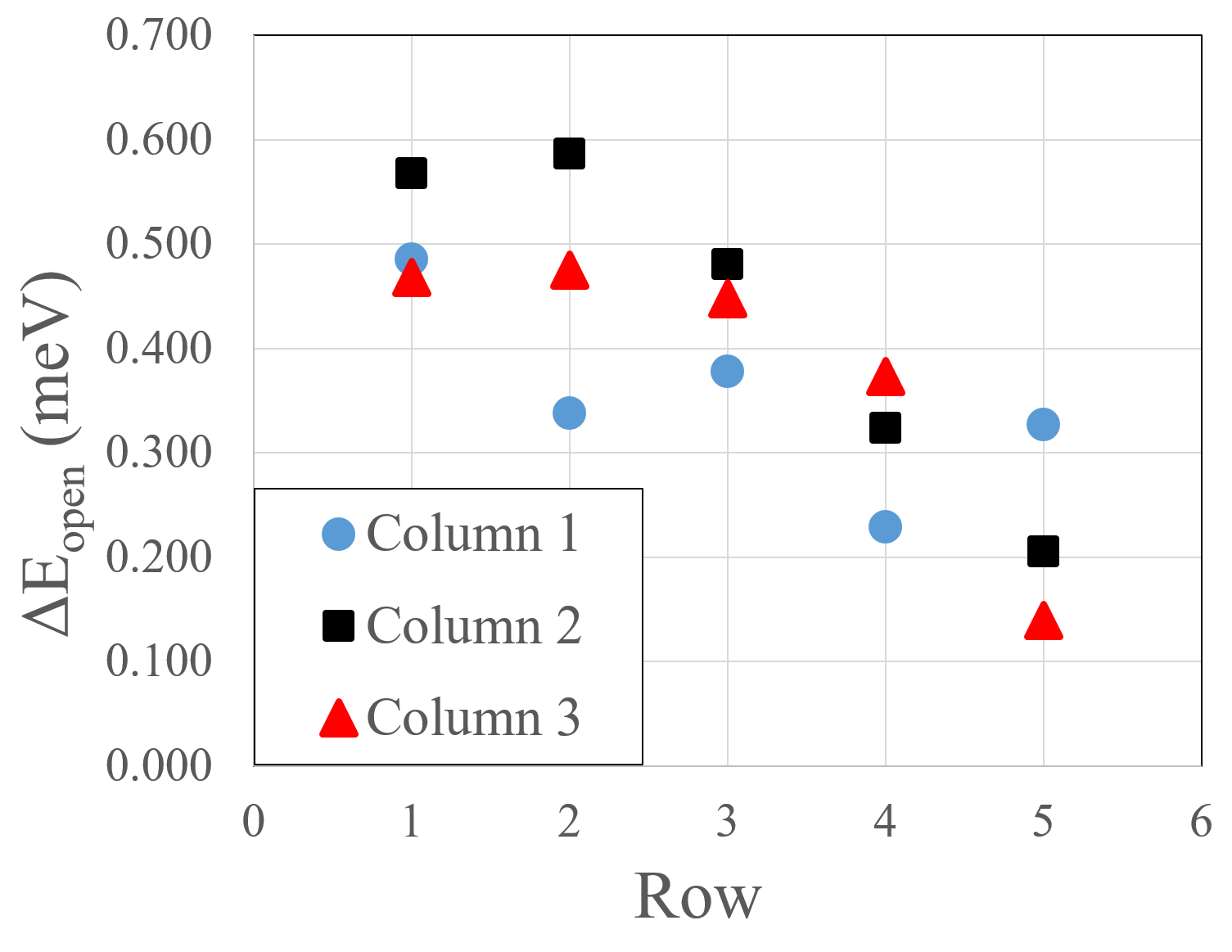}   
	\caption{{\bf{Gate dependence}} Dependence of $\Delta E_{\rm{open}}$ on $V_{\rm{LGS}}$ for each column of the honeycomb in figure~\ref{grid}. Each data point shown is an average of all the $\Delta E_{{\rm{open}}}$ determinations for a given transition.  Some of these averages include data taken with source and drain connections reversed as a check against any extrinsic voltage offsets.  There is a general trend of decreasing $\Delta E_{\rm{open}}$ as the source dot is successively filled. This trend comes about naturally in the valley band model (see text). \label{gatedependence}}
\end{figure}

Size asymmetries of this nature are typically associated with current rectification due to a metastable excited state of the electrons on the DQD. The most common example is PSB\cite{Nadj10,Johnson05B,Lai11,Yamahata12,Simmons10,Koppens05,Shaji08,Churchill09A,Weber14,Liu08,Pfund07}, where the occupation of a spin triplet state in the (1,1) charge configuration prevents current flow that would otherwise be allowed through the ground singlet states\footnote{In the usual PSB model the electron occupations of the dots are considered to be effectively (1,1) and (0,2), and any other electrons on the dot are forming closed shells in lower orbits and thus inert. Throughout this manuscript when we reference the (1,1) and (0,2) charge states we are referring to effective charge states like these.}.  Comparisons between our data and qualitative expectations of PSB reveals several inconsistent features.  First, the size asymmetry of figure~\ref{triangles} exists at nearly all the transitions shown in figure~\ref{grid} whereas  PSB, generically being an odd-even filling effect, is expected in, at most, 1/2 of the transitions\footnote{PSB in silicon has been reported with unusual fillings, most notably in reference~[\onlinecite{Yamahata12}]. These unusual electron fillings are typically explained using the valley degree of freedom in silicon, which allows for higher spin state formation\cite{Hada03} which could allow PSB to occur at consecutive transitions.} . Second, all of the asymmetries observed had the same polarity (larger triangles measured with $V_b > 0$). Both of these observations are shown in figure~\ref{gatedependence}, which has the values of $\Delta E_{open}$ for all 15 transitions shown in figure~\ref{grid}. In contrast, PSB is expected to show size asymmetries with alternating polarity as one moves through the honeycomb\cite{Johnson05B}. Third, as shown in figure~\ref{Bfield}, our data does not have a systematic trend with respect to magnetic field $B_0$ applied perpendicular to the substrate. Although the magnetic field does change the magnitude of the size asymmetry the dependence is not what is expected from simple spin states as one might expect with PSB. In PSB, one expects a systematic change in the size of the bias triangles due to two effects. 1) the exchange energy can have a magnetic field dependence\cite{Johnson05B}, and 2) the energy of the polarized spin triplet states have a magnetic field dependence due to the Zeeman effect\cite{Lai11,Perron16}. Fourth, as shown in figure~\ref{gatedependence}, our data show a systematic dependence on $V_{\rm{LGS}}$ that is unexpected for the case of PSB. Although changing the voltage applied to a barrier gate can change the magnitude of the exchange energy\cite{Koppens05}, the effect is too small to be a plausible explanation for the trend we observe. 

\begin{figure*}
	\centering
	\includegraphics[width = 0.9\textwidth]{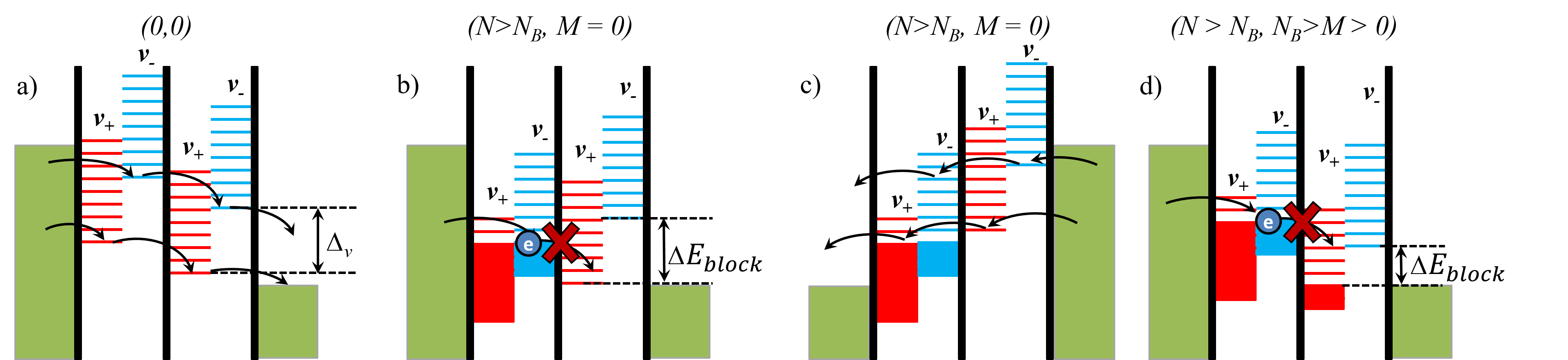}   
	\caption{{\bf{Model:}} Our model in the simplified constant interaction picture. We have suppressed the charging energy to evoke the idea of bands. {\bf{a)}} Red and blue levels correspond to $v_+$ and $v_-$ valley types respectively. When the dots are empty $(0,0)$, the ground states in each dot are the same valley type, and the splitting to the lowest state in the opposite valley is also equal. This results in no blockade, and bias triangles with size determined by $|V_b|$. {\bf{b)}} Filling the first dot with $N>N_B$ electrons removes the splitting between the chemical potentials of the two valley types. Blockade will occur when these degenerate levels lie between the ground $v_+$ and $v_-$ states of the second dot, resulting in a bias triangle size asymmetry corresponding to $\Delta E_{\rm{block}}$. {\bf{c)}} Reversing the bias polarity in this situation results in fully formed bias triangles with no blockade. Whenever the interdot tunneling is favorable for the lower $v_+$ states the same is true for the $v_-$ states. {\bf{d)}} Adding M electrons to the second dot reduces $E_{\rm{block}}$, and thus the size asymmetry, by filling $v_+$ states and bringing the $v_+$ chemical potential closer to that of $v_-$ leading to the dependence seen in figure~\ref{gatedependence}. \label{model}}
\end{figure*}

The inconsistencies between our data and traditional PSB model lead us to believe the asymmetries in our bias triangles are not due to simple spin states; therefore, another degree of freedom must contribute to the rectification.  We developed a model of blockade that centers on the valley degree of freedom of electrons in silicon. In our model, we assume all the conduction electrons in the DQD occupy one of two valley states, $v_+$ and $v_-$. However, the relative large size and electron occupation of our dots results in bands of states for each valley type (figure~\ref{model}).  The bottom of the bands of each type are separated by the valley splitting $\Delta_v$, which depends primarily on the surface potential experienced by the electrons\cite{Culcer10B}. The splittings between successive levels in the individual bands are predominantly determined by the orbital spacing $E_{\rm{orb}}$. 

Figure~\ref{model} shows the details of this model in the simplified constant interaction picture where we have suppressed the charging energy to evoke the idea of bands, appropriate near the triple points of the honeycomb. In the limit of empty dots the ground state chemical potential is the same for both dots. This situation results in typical bias triangles with no asymmetry.  However, by adding electrons to one of the dots we fill the lower lying valley states.  At some number, $N_B$, of electrons all the $V_+$ states below the bottom of the $V_-$ band will be filled and the chemical potenial for the two valley types will be degenerate on that dot. With $M$, the number of electrons on the other dot, being less than $N_B$ the chemical potentials for the two valley types are split on the second dot.  This situation, one dot with degenerate chemical potentials for the two valley types and one with split chemical potentials, allows for a blockade to occur. An electron loads into a $v_-$ state on the first dot only to become trapped since the interdot tunneling event is energetically unfavorable for $v_-$ states. Thus, current is blocked until some valley relaxation or inter-valley tunneling event occurs\footnote{This argument assumes that current cannot flow by creating a hole in the other valley state on the left dot; however, we expect this process to be suppressed because the state with an inter-valley electron-hole pair on the left dot is an excited state of the dot, which will be gapped from the ground state due to finite-size effects.}. This situation is shown in figure~\ref{model}b for $(N>N_b,M=0)$ and figure~\ref{model}d for $(N>N_B, N_B>M>0)$. However, when the bias is reversed, the second dot in the conduction path now has degenerate chemical potentials for the two valley types and no blockade occurs (see panel c).  As with PSB, this blockade results in a bias triangle size asymmetry $\Delta E_{\rm{open}}$. The magnitude of $\Delta E_{open}$ corresponding to $E_{\rm{block}}$, the energy splitting between the lowest unoccupied $V_-$ state and the highest occupied $V_+$ state on the second dot. $E_{\rm{block}}$ can be reduced by adding electrons to the blocking dot and successively filling the lower energy valley band as shown in panel d).

Applying this model to our system we see it predicts a blockade that is in qualitative agreement with our data. A drain-side dot filled such that $N>N_B$ and a source-side dot in the $N_B>M>0$ regime would lead to the size asymmetry we observe. Specifically, 1) blockade for multiple successive transitions and 2) the same polarity of size asymmetry for these transitions.  Furthermore, the magnetic field dependence of the bias triangle size would depend heavily on the details of the states that make up the bands, and is by no means expected to be systematic or monotonic.  Finally,  adding electrons to the source-side dot would reduce $E_{\rm{block}}$ and therefore reduce $\Delta E_{open}$. In our data this would correspond to moving vertically through the honeycomb in figure~\ref{grid} and results in the reduction in $\Delta E_{\rm{open}}$ seen in figure~\ref{gatedependence}. 

A crucial assumption is that the valley degree of freedom is a good quantum number; the symmetric and anti-symmetric valley states represented in figure~\ref{model} are eigenstates of the combined Si band structure/interface, and in the absence of large interface roughness\cite{Huang17}, the valley states represent a good quantum number. Furthermore, for the model to apply several things must be true. First, there must be low inter-valley tunneling rates. Furthermore, the $z$ orbital spacing $E_{\rm{z}}$, where $z$ is the direction perpendicular to the substrate, must be large relative to both $\Delta_v$ and the lateral orbital spacing $E_{\rm{orb}}$. If this were not the case, adding electrons to one of the dots would cause significant changes to the $z$-dependence of the wave function, which would distort the relative valley states on each dot, leading to inter-valley tunneling. This requirement seems probable in our device where the lithographic distance between barrier gates is 40~nm, while a typical thickness for a silicon MOSFET inversion layer is roughly a tenth of that. In addition to the restriction on $E_{\rm{z}}$, the lateral orbital spacing must be small relative to $\Delta_v$ for the band-like picture to be accurate. Using a constant interaction picture and applying this model to our data we can extract certain energies. The largest splitting in figure~\ref{gatedependence} of $\sim$0.6~meV gives a lower bound on $\Delta_v$. This value is in the range of 0.1~meV to 0.8~meV reported by references [\onlinecite{Yang13}] and [\onlinecite{Goswami07}]. The slope of figure~\ref{gatedependence} implies a lateral orbital splitting $E_{\rm{lat}} \sim 0.1$~meV.  Furthermore, the noise level in our experiment of $\pm 0.1$~pA implies a valley lifetime of $T_{1,v} \geq 1~\mu$s. 


The model we have described, although consistent with our data,  needs to be verified with future measurements.  Devices with extended functionality will allow for more quantitative comparisons between the model and experiment. Specifically, independent plunger gates for each of the dots as well as an ancillary dot for charge sensing would allow for the exact electron fillings to be determined. This ability would allow one to confirm the differences in electron occupation and examine other regions of the charge stability diagram where the model predicts the same type of blockade but in the opposite direction. Charge sensing would also allow for a much wider range of state lifetimes to be probed. In addition to verifying the model, this would open the door to investigations of the coupling and relaxation mechanisms\cite{Johnson05A}.  

The authors would like to thank Josh Pomeroy and Garnett Bryant for their fruitful discussions as well as Akira Fujiwara for providing devices. 


%
\end{document}